# Application of Transformer Impedance Correction Tables in Power Flow Studies

Pooria Dehghanian, Ju Hee Yeo, Jessica Wert, Hanyue Li, Komal Shetye, Thomas J. Overbye
Department of Electrical and Computer Engineering, Texas A&M University, College Station, Texas, USA
{pooria.dehghanian; yeochee26; jwert; hanyueli; shetye; Overbye}@tamu.edu

*Abstract*— **Phase Shifting Transformers (PST) are used to control or block certain flows of real power through phase angle regulation across the device. Its functionality is crucial to special situations such as eliminating loop flow through an area and balancing real power flow between parallel paths. Impedance correction tables are used to model that the impedance of phase shifting transformers often vary as a function of their phase angle shift. The focus of this paper is to consider the modeling errors if the impact of this changing impedance is ignored. The simulations are tested through different scenarios using a 37-bus test case and a 10,000-bus synthetic power grid. The results verify the important role of impedance correction factor to get more accurate and optimal power solutions.**

## I. Nomenclature

| | |
|---|---|
| $I_i$ | Injection current at bus $i$ |
| $i_i$ | Outgoing flow current at bus $i$ |
| $I_k$ | Injection current at bus $k$ |
| $i_k$ | Outgoing flow current at bus $k$ |
| $V_i$ | Bus voltage at bus $i$ |
| $V_k$ | Bus voltage at bus $k$ |
| $T$ | Off-nominal turns ratio of the PST |
| $E_i$ | Transmission line voltage |
| $y$ | Admittance of the PST |
| $\theta_{ik}$ | Angle difference between the buss's voltage |
| $Z_{line}$ | Transmission line impedance value |
| $\phi$ | Phase shift of the transformer between bus $i$ to $k$ |
| $Z_T$ | Impedance of transformer with no consideration of correction factor |
| $K_T$ | Correction factor index |
| $Z_{TK}$ | Impedance of transformer considering correction factor |
| $V_{rT}$ | Rated voltage of the transformer |
| $S_{rT}$ | Rated power of transformer |
| $x_T$ | Reactance of transformer (per unit) |
| $c_{max}$ | Voltage factor corresponding to low voltage side of the transformer |

## II. Introduction

TRANSFORMERS are considered as one of the key elements in power grids which enable a safe and efficient transfer of power supply to the end user. Power flow control and management tend to be challenging due to uncertainty of transmission system operation caused by integration of various control devices, incremental increase in renewable energy sources, and high load penetration into the power system. Because of complexity of the transmission network together with variability and deployment of renewable resources located distantly from the demand centers, system operators may face congestions in transmission power lines which may threaten the security and stability of the entire system. Congestion management plays a vital role in building an efficient energy market operation. To alleviate the congestion, transmission system operators may apply possible solutions such as generation re-dispatch, network reconfiguration, counter trading and transaction curtailment [1].

To meet the variety of requirements defined in NERC standards, Phase Shifting Transformers (PSTs) maybe a useful solution. PSTs, which are sometimes referred to as "*phase angle regulators*", enhance operational and economic benefits by controlling real power flows, both in magnitude and direction, across transmission lines by phase angle manipulation [2]. A PST forces a voltage's phase shift (±90°) using its tap changer between the primary (source) and secondary (load) terminals, which changes the branch impedance according to the off-nominal phase shift value of the PST [3]. This is where the need for impedance correction tables arises. Change of the phase angle also changes the transformer's impedance which makes the Y-bus matrix asymmetric, leading to a more complicated power flow solution calculation [4]-[7].

PST can be categorized based on their characteristics. They could be single or dual core, symmetric or asymmetric design, and quadrative or non-quadrative type. PSTs are a type of flexible ac transmission system (FACTS) device with variable-impedance features which may have power rating higher than 1000 MVA [8]. A PST's main purpose is to regulate the MW flow in the transmission grid. An important configuration of a PST is when it is connected to one line to control the flow of the parallel line and avoid loop congestion [9]. Additionally, it reduces the line losses and thus, increases the grid capacity and security. Transformer's impedance winding calculation at each tap position is necessary to get a reliable PST model for both a balanced and unbalanced system operation. These calculations are usually made through zero, positive, and negative sequence models [10]. Typically, transformer manufactures provide detailed data sets including the results of open-circuit and short-circuit tests, design parameters, impedance correction multipliers, positive and zero sequence impedance, and maximum no-load phase shift, among others.

In this paper, the impact of applying the impedance correction (multiplier) factor into the PST setting on power flow study, especially in congestion mitigation and contingency analysis, is highlighted through testing various scenarios. If the impedance correction factor is applied into the transformer data set, it changes the value of the impedance of transformer.

The total impedance of the transformer is adjusted by this factor using a piece-wise linear curve. This causes the corresponding elements in the bus admittance matrix to change, complicating the power flow solution but increasing its accuracy. The paper investigates how impedance correction table consideration impacts PST operation during various system operating conditions.

More specifically, the contribution of the paper is to emphasize the possible modeling error introduced when impedance correction tables are not considered in power flow study.

The general role of PSTs has been considered in a number of prior papers. For example, the installation of PSTs in the actual Indian grid with the aim of relieving overloaded lines is justified in [11]. The role of PST in loop flow mitigation is mainly investigated in [12]. One possible placement of a PST is in the cross-borders in case of circular flows to ensure stability of power supply. This approach has been utilized as a short-term solution in certain crucial points of the European grid such as in Netherlands, Poland, and Belgium [13]. A fast-decoupled power flow method in presence of asymmetric PST is proposed in [14]. The authors analyzed the eigenvalues of the matrix to find the features of the power flow deviation caused by PST. In another study [15], the impact of utilizing PSTs on minimizing cost of remedial actions of transmission grid is investigated. Also, it is claimed that the system losses may rise due to operation of PST since the power flow can be routed through higher impedance lines on longer distances. In [16], a line flow sensitivity analysis in the presence of PST is conducted to assess the role of PST in power flow analysis. A modified PST model is integrated into the standard direct approach power flow formulation in [17] to investigate its impact on the power flow convergence. The influence of application of PST in parallel EHV interconnection transmission lines on voltage stability is addressed in [18].

However, few studies have been conducted to show how the power flow solution changes when the impedance correction data sets are applied to the PSTs. The importance of using impedance correction table in PST to create a more realistic synthetic power grid is discussed in [3], [19]. In [19], the role of different impedance tables on power flow solutions are investigated. This paper builds on [19] by focusing on the applications of impedance correction tables; and the modeling error that would be introduced if impedance corrections are ignored.

The rest of the paper is organized is as follows:
Section III introduces the formulation of impedance correction factor into PST and its impact on power loss. Section IV includes the simulation and numerical results of the case studies. Finally, the conclusion is presented in Section V.

## III. PROBLEM FORMULATION

### A. Phase Shifting Transformer Model

An equivalent circuit of a PST is shown in Fig.1 [19]. Note that the line admittance is in series with the PST. The amount of real power flow of the transmission line connected between buses $i$ and $k$ can be obtained as follows:

$$P = \frac{|V_i| \cdot |V_k|}{Z_{line}} \sin \theta_{ik} \quad (1)$$

Equations (2) and (3) express the relationship between voltage and current of an ideal transformer and phase shifting transformers, respectively [6].

$$\frac{V_i}{E_i} = T\angle\phi = Te^{j\phi} = T(\cos\phi + j\sin\phi) = \alpha + j\beta \quad (2)$$

$$\begin{bmatrix} I_i \\ I_k \end{bmatrix} = \begin{pmatrix} \dfrac{y}{\alpha^2 + j\beta^2} & -\dfrac{y}{\alpha + j\beta} \\ -\dfrac{y}{\alpha + j\beta} & y \end{pmatrix} \begin{bmatrix} V_i \\ V_k \end{bmatrix} \quad (3)$$

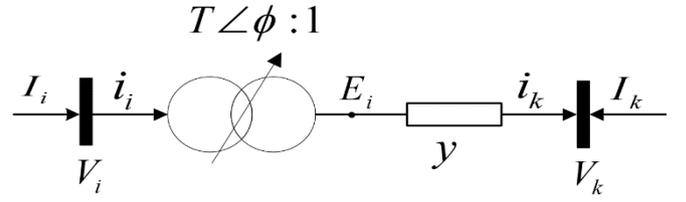

Figure 1. Single-line equivalent circuit of phase shifting transformer.

As we can see from (3), the admittance matrix is asymmetric i.e. the transfer admittance between two buses is not equal.

### B. Impedance Correction Factor Calculation

As previously discussed, integrating the impedance correction table changes the transformer impedance by manipulating the phase angle. Based on the IEC 60909-0 guideline [20], the impedance correction factor $K_T$ can be calculated using (4)-(5) as follows:

$$Z_T = R_T + jX_T \quad (4\text{-a})$$

$$Z_{TK} = K_T * Z_T \quad (4\text{-b})$$

$$x_T = X_T * \frac{S_{rT}}{V_{rT}^2} \quad (4\text{-c})$$

$$K_T = 0.95 * \frac{c_{max}}{1 + 0.6\, x_T} \quad (5)$$

### C. Power Loss Calculation Considering Impedance Correction Factor

A correction table also can be used to properly model the impedance of the transformers. Applying the impedance correction factor to the PST's model changes the value of impedance of the transformer by regulating the phase angle. Therefore, resistance of a power line which connected to a PST can be adjusted as presented in (6).

$$P_{Loss,\phi} = R \times \{K_T(\phi)e^{j\phi}\} \times I^2 \quad (6)$$

where $\phi$ is the phase angle of the PST and $K_T(\phi)$ is the correction factor at a specific angle. Using a correction table can get a more accurate assessment of system losses and make good planning decisions to protect the system from overloads, loop flows etc.

## IV. CASE STUDY AND SIMULATION RESULTS

The main contribution of this paper is to consider the potential modeling error that would be introduced if the role of impedance correction table is ignored in power flow study. The following case studies are presented to verify the possible impact of modeling the impedance correction table into the PST data sets on power flow control, especially within the PST's local footprint. The effect of impedance correction tables on power flow solutions is assessed through metrics such as: MW line flow changes, voltage violation reductions, and Available Transfer Capability (ATC) threshold.

### A. Case Study I: 37-Bus test System

The 37-bus system is a publicly available test case [21]. The total load demand and generation capacity are 1596.9 MW and 1635.3 MW, respectively. This system has one 350/69 kV PST. The stressed test system is created to assess if PST with impedance correction factor can reduce power losses and alleviate stressed situations. The one-line diagram of the stressed 37-bus system is shown in Fig. 2. Usually manufacturers provide different impedance correction tables depending on the characteristic design of a PST. There are a number of different tables but, in this paper, the impedance correction factor, which operates as a function of phase angle, presented in Table I (as indicated in [24]) is implemented into the PST. The nominal impedance value of the PST is multiplied by the impedance correction (scale) factor, $K_T(\phi)$, to get the actual value. A piecewise linear interpolation method is used to determine the appropriate scaling factor base on the phase angle. The total MW losses, bus voltage magnitude, and line flow capacity are observed, and the results are compared in two different cases: a) PST without impedance correction table; b) PST with impedance correction factor tabulated in Table I.

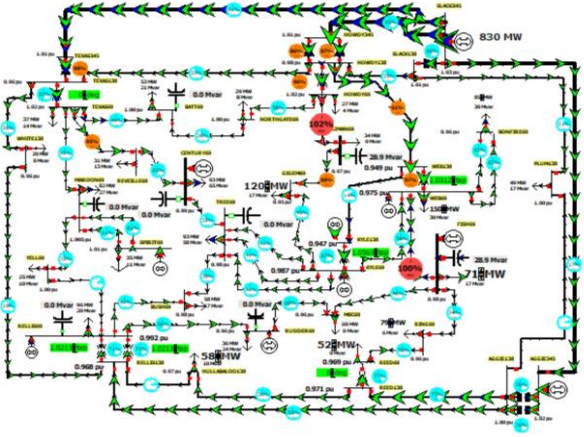

Figure 2. One-line Diagram of 37-bus Stressed System.

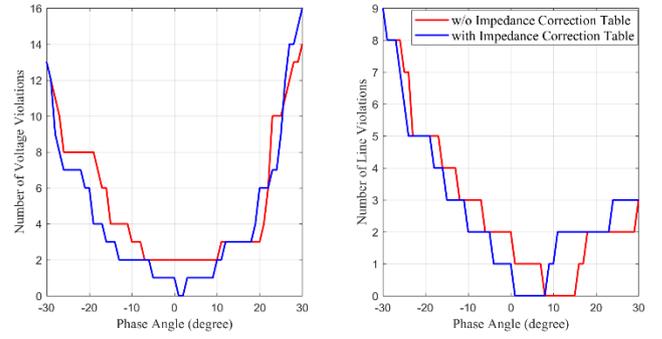

Figure 4. Impact of correction table consideration on the line and voltage violations profile.

Fig. 3 illustrates the total system MW losses. The amount of MW loss changes with the phase angle variation. The results reveal that utilizing the correction factors can decrease the power loss more efficiently than when they are ignored. This highlights the fact that impedance correction table changes the losses. Although the losses might easily go up for different cases, the changes are what is important. Fig. 4 shows that the impedance correction table can also aid in alleviating stressed situations in the system and providing a more accurate power flow solution.

TABLE I
THE UTILIZED IMPEDANCE CORRECTION TABLE

| $\phi$ | -152 | -121 | -85 | -42 | 0 |
|---|---|---|---|---|---|
| $K_T(\phi)$ | 1 | 0.62 | 0.37 | 0.21 | 0.15 |
| $\phi$ | 42 | 85 | 121 | 152 | |
| $K_T(\phi)$ | 0.21 | 0.37 | 0.62 | 1 | |

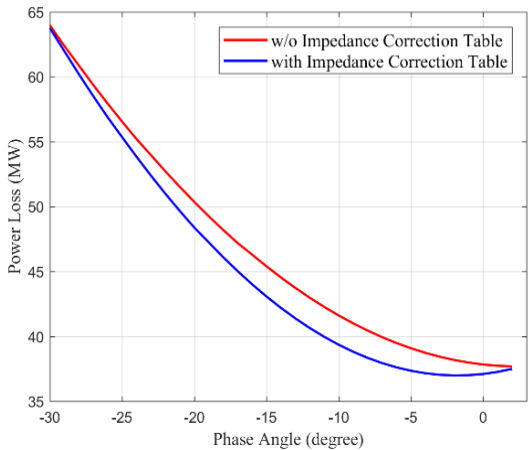

Figure 3. The total MW losses observed in 37-bus system.

The voltage violation limit is when a bus voltage is less than 0.95 (p.u.) or greater than 1.05 (p.u.). The stressed base case system already has two low voltage issues at bus 47 and bus 53 when the phase angle is 0°. The results show that the voltage violations are not reduced when impedance correction factors are ignored. Instead, more violations occur with changes in the phase angle. On the other hand, using correction factors resolves the low voltage issues by regulating its angle to 1° or 2°. For example, the voltage magnitudes at bus 47 and bus 53 become 0.951 and 0.95 p.u., respectively when the phase angle is 1°. For line violations, the MVA flow from at least one branch must be over 100% of the limit. The base case has two overloaded lines, i.e. branch 10-13 and branch 20-48, at phase angle $\phi = 0°$. Figure 4 shows that these overloading issues are cleared up at 1° and 2° with the impedance correction table. Note that, in this case, both overloading and low voltage issues are resolved simultaneously utilizing the impedance correction table.

### B. Case Study II: 10,000-Bus Synthetic Power Grid

A totally fictious 10,000 bus synthetic power grid [22], [23] geographically based over the footprint of western United States is used to address the role of impedance correction data sets of a PST in alleviating the system instability and contingencies analysis. Different scenarios are created and tested to investigate the impact of impedance correction table on MW line flow violations, voltage issues, and other scheduled transactions between different areas of the system. The base case system contains 10,000 buses, 9726 transmission lines, 2485 generating units with a total capacity of 153,396 MW serving 150,917 MW of load among 4899 load points. Fig. 5 illustrates the per unit bus voltage contour using a one line diagram of the 10k-bus case and the line flows (i.e., transmission lines and transformers) and percentage loads are showing using flow arrows and pie charts, respectively [25], [26]. (Note: this system is generated from public data and does not correspond to actual grid.)

> Scenario 1: Power Flow Control Considering the Impedance Correction Table Sets of Phase Shifting Transformers

In this simulation we demonstrate that ignoring the implementation of impedance correction table into the PST setting may lead to an incorrect optimal power flow. Motivated by the 10,000-bus system shown in Fig. 5, a strongly stressed test case is created to better highlight the influence of impedance table on line flow. Some additional modifications have been made to direct more flows from the Washington and Oregon areas into Northern California (CA). This causes Northern and Central CA areas to experience a heavy stress mode. In doing so, the load in CA is increased by 20% in the North/Central and Bay area. The generation in CA is decreased by 7% to prevent local generators from supplying the load increase.

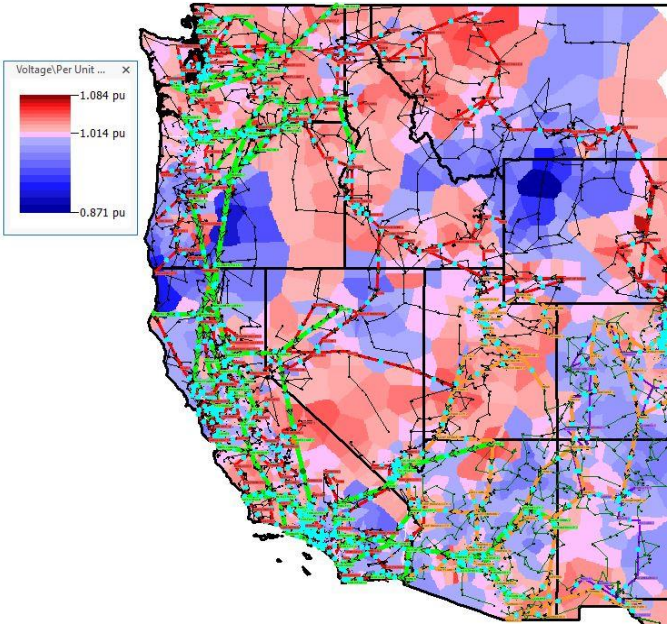

Figure 5. One-line diagram of 10,000-bus synthetic power case- Bus voltage contour.

Also, the generation in Arizona area is reduced by 10%, followed by a 765 kV line outage connecting Southern and Northern CA area to prevent generation in Arizona and Colorado areas ramping up to supply the load increase.

The above stressed scenario yields overloaded line and low bus-voltage issues in the test case. The goal is to compare the role of PSTs, a) with and b) without impedance correction tables considerations, in mitigating the violations. This approach helps us to realize how impedance correction table consideration provides a more accurate modeling for PST application. In this simulation, the line connected between buses 23488 and 23413 exceeds its MVA capacity by 102%.

Fig. 6 zooms into the congestion zone in Central CA area. Considering the topology and loadability of the system, and through looking up the locations of the PSTs near the congestion area, a PST is selected, which is connected to the bus 23414 and 23413 and roughly lies in the congestion zone. Table II shows the PST performance (with and without impedance correction) in handling the overloaded line. For both cases, incrementally increasing the angle towards the negative direction increases the MVA flow on the overloaded line. However, increasing the angle in the positive direction leads to possible mitigation. At angle $0°$, using the impedance correction table reduces the line flow by 4% compared to the base case where no impedance correction table does not solve the excessive flow of the line. The same trend can be seen for the angle $2°$.

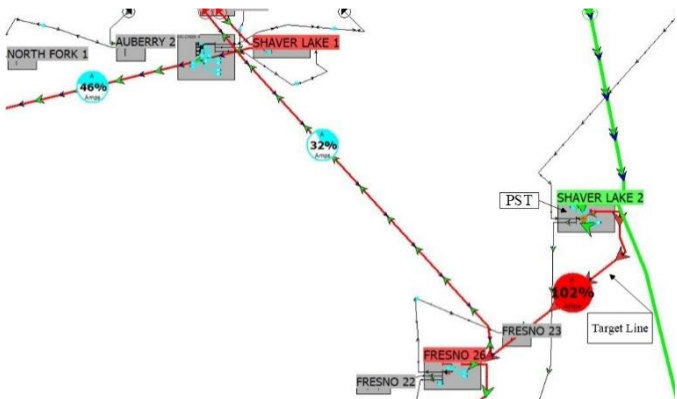

Figure 6. Zooming into the Congestion area. The PST is located in Shaver Lake 2 Substation.

TABLE II
IMPACT OF CONSIDERING IMPEDANCE CORRECTION TABLE SET ON FLOW BETWEEN BRANCH 23488-23413

| PST without Impedance Correction Consideration | | | | PST with Impedance Correction Consideration | | | |
|---|---|---|---|---|---|---|---|
| $\phi$ | $P_\phi$ (MW) | MVA flow (%) | Extra Line violation | $\phi$ | $P_\phi$ (MW) | MVA flow (%) | Extra Line violation |
| -18 | 1472 | 120 | - | -18 | 1492 | 121 | - |
| -10 | 1376 | 112 | - | -10 | 1369 | 111 | - |
| -4 | 1303 | 106 | - | -4 | 1273 | 103 | - |
| 0 | 1255 | **102** | - | 0 | 1209 | **98** | - |
| 2 | 1230 | **100** | - | 2 | 1177 | **96** | - |
| 6 | 1182 | 96 | - | 6 | 1114 | 91 | - |
| 10 | 1037 | 93 | - | 10 | 1054 | 86 | - |
| 14 | 1090 | 89 | - | 14 | 994 | 81 | 2 |

Note: $\phi$ is the voltage angle of PST; $P_\phi$ is the MW power flow through the overloading line; Base MVA(%) is the line power rate.
-Extra line violation means the other overloading line following the change of angle. (the primary overloaded target line is not counted)

Still increasing positive angle yields more line flow percentage reduction when impedance table is considered. However, at angle $14°$, with impedance correction, the target line not overloaded but two other lines overpass their capacities where one is the line that is connected to the PST itself. So, the maximum positive angle in this case should be less than $-14°$. The other overloaded line is roughly near the congestion area as expected. Therefore, the results verify that at least within local areas utilizing impedance correction table can even lead to a better optimal solution.

➢ *Scenario 2: Assessing voltage violation in presence of the Impedance Correction Table*

One of the reasons that we have created a fully stressed test case is to mimic an actual grid and investigate some aspects of power system operation, here voltage violation, when taking the impedance correction table set into consideration. In this scenario (S2), the same stressed 10,000-bus case as (S1) is used to further analyze whether the triggered low voltage area in Northern CA can be improved using impedance correction table.

Due to an imbalanced generation and load in Northern CA the power grid faces low voltage at four bus stations (b_20000, b_20004, b_20008, and b_20009). Fig. 5 demonstrates the bus voltage contour of the test system. Note that at the beginning with no consideration of impedance correction table, all four busses have the per unit minimum limit voltage lower than 0.9 V (p.u.). The total amount of load and generation in both CA and OR areas is tabulated in Table III.

To address the impact of correction table on the four low-voltage busses two PSTs are selected. PST_1 is between busses 13058-13059 relatively closer to low-voltage zone and PST_2 is placed between busses 13548-13549. The simulation shows that although the PST_2 is relatively farther to the four low-voltage busses, it has greater impacts on alleviating the low-voltage issue than the PST_1.

TABLE III
LOAD AND GENERATION LEVEL OF THE LOW-VOLTAGE AREAS UNDER STUDY

| Area | Load | | Generation | |
|---|---|---|---|---|
| | MW | Mvar | MW | Mvar |
| Northern California | 7239 | 1805 | 7094 | 2509 |
| Oregon | 9277 | 24.61 | 12580 | 1057 |

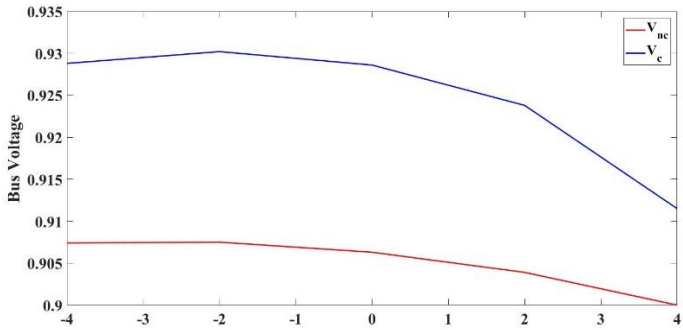
Fig. 7(a). Change of voltage at bus 20000 to different phase angle.

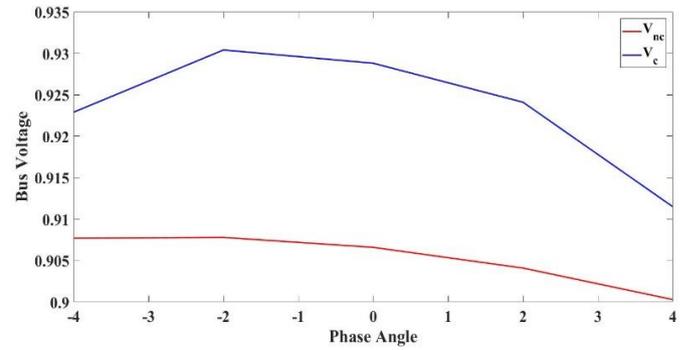
Fig. 7(b). Change of voltage at bus 20004 to different phase angle.

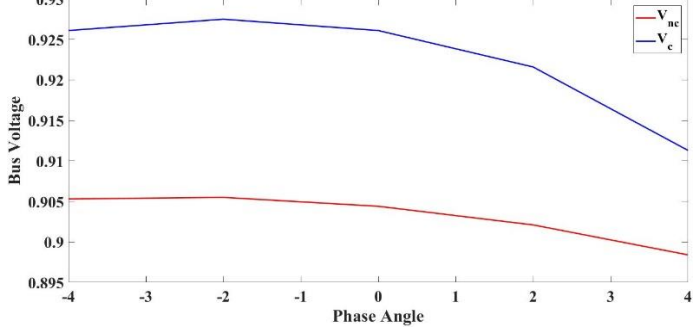
Fig. 7(c). Change of voltage at bus 20008 to different phase angle.

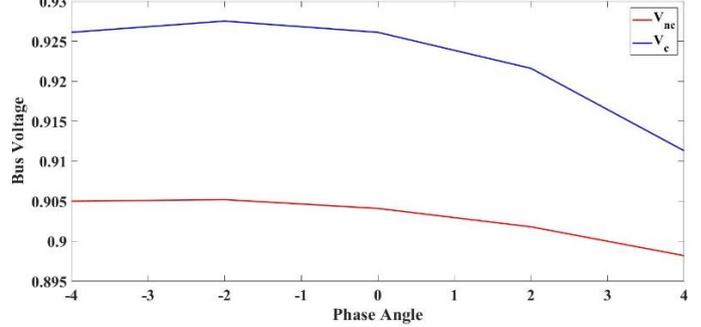
Fig. 7(d). Change of voltage at bus 20009 to different phase angle.

Figure 7. Impact of PST_2 with and without impedance correction table consideration on bus voltage profile.

TABLE IV
BUS VOLTAGE PROFILE CHANGES DUE TO PST_1

| $\phi$ | Bus20000 | | Bus20004 | | Bus20008 | | Bus20009 | |
|---|---|---|---|---|---|---|---|---|
| | Vc | Vnc | Vc | Vnc | Vc | Vnc | Vc | Vnc |
| +4 | 0.894 | 0.890 | 0.895 | 0.890 | 0.892 | 0.888 | 0.891 | 0.888 |
| +2 | 0.904 | 0.894 | 0.904 | 0.894 | 0.901 | 0.892 | 0.901 | 0.892 |
| 0 | 0.908 | 0.896 | 0.908 | 0.896 | 0.905 | 0.894 | 0.905 | 0.894 |
| -2 | 0.907 | 0.901 | 0.907 | 0.901 | 0.905 | 0.899 | 0.905 | 0.899 |
| -4 | 0.901 | 0.897 | 0.902 | 0.898 | 0.900 | 0.896 | 0.900 | 0.896 |

Figs. 7(a-d) illustrate the low voltage-bus enhancement when the impedance correction table is set into the PST_2. Table IV presents the bus voltage profile due to change of phase angle of PST_1 with and without correction table. Note that in both Fig. 7 and Table IV, $V_c$ and $V_{nc}$ denote the bus voltage profile with and with no impedance correction table consideration, respectively. Table IV and Fig. 7 show that the voltage profile improved when the impedance correction is considered but the role of PST_2 on voltage enhancement is greater than PST_1.

Generally speaking, PSTs are not used for the purpose of maintaining the voltage issues as they mainly change the active power of the line rather than the reactive power. However, the simulations verify that impedance correction table also could play a positive role in improvement of the voltage profile.

➢ *Scenario 3: ATC Control through Impedance Correction Factor Consideration*

ATC is the maximum available additional power that can be transferred from area A to B on top of the power that already committed in use and without compromising the system security. ATC plays a great role especially when power sellers and buyers share a common transmission network [27]. Therefore, accurate determination of the ATC is of great importance for the purpose of economic benefit and system operation security. ATC can be calculated using linear methods in DC power flow.

In this scenario, the impact of the inclusion of impedance correction tables in PST models is assessed under conservative and moderate PST integration conditions. In each, the buyer area is defined to be Northern California and the seller consists of all generators outside of the Northern California area. ATC is performed considering PSTs with and without impedance correction tables. The impedance correction table used is presented in Table II.

- *Condition A: Conservative use of PSTs*

Under Condition A, three PSTs are used. One is placed in Northern California (bus 20006 to bus 20005), Oregon (bus 13529 to 13548), and Washington (bus 10784 to 10788).

Including the impedance correction table for the PSTs results in a calculated ATC of 190.0 MW. When excluding the impedance correction table from the PST model, the ATC is 119.9 MW. Thus, the results show that including impedance correction factor changes the solution whether the PST with impedance correction table has a greater impact on ATC value.

- *Condition B: Moderate use of PSTs*

Under Condition B, nine PSTs are placed strategically throughout the 10,000-bus system. These PSTs are located in the following areas: Central California (bus 23414 to 23413), Montana (bus 77254 to 77262 and bus 77254 to 77262), Northern California (bus 20006 to bus 20005), Oregon (bus 13529 to 13548), Southeast California (bus 28737 to 28745), Southwest California (bus 25870 to 25869), Utah (bus 50203 to 50297), and Washington (bus 10784 to 10788). Including the impedance correction table for the PSTs yielded a calculated ATC of 732.8 MW. When the modeling of the PSTs excludes the impedance correction table, the calculated ATC is 162.8 MW. The discrepancy between these results, based on the modeling choice to include or exclude impedance correction tables, is 570 MW in this scenario.

In both scenarios, including impedance correction table yields a higher ATC value for a transaction between Northern CA to the rest of the system. However, the changes are what is more important regardless of increase or decrease in power flow solutions if impedance table is utilized and, hence, impedance scaling factor should be considered in power flow simulations and modeling.

## V. Conclusion

The focus of the paper is to consider the modeling error introduced in phase shifting transformer setting when impedance correction table set is ignored. To evaluate the impact of impedance correction factor consideration on power flow study some indices are selected such as MW line power loss, bus voltage and ATC limit. The simulations are tested in a 37-bus test system and further the influence of impedance correction factor is validated through a highly stressed 10,000-bus synthetic grid. Several scenarios are created to better highlight the role of impedance correction factor on mitigating MW overload line, stabilizing bus voltage profile, and maximizing MW ATC between areas. The simulated results verify that considering embedding impedance correction factor into PST can lead to a more optimal solution especially in the local zone of contingencies and should be greatly considered in power flow analysis and simulation tools.

## Acknowledgement

The work presented in this paper funded in part by the US Department of Energy Advanced Research Projects Agency-Energy (ARPA-E). The authors gratefully acknowledge this support.

## References

[1] N. I. Yusoff, A. A. M. Zin and A. Bin Khairuddin, "Congestion management in power system: A review," *3rd International Conference on Power Generation Systems and Renewable Energy Technologies (PGSRET)*, Johor Bahru, 2017, pp. 22-27

[2] J. Boudrias, "Power factor correction and energy saving with propertransformer, phase shifting techniques and harmonic mitigation," *LargeEngineering Systems Conference on Power Engineering (IEEE Cat.No.04EX819)*, pp. 98–101, 2004.

[3] A. B. Birchfield, T. Xu, and T. J. Overbye, "Power flow convergence and reactive power planning in the creation of large synthetic grids," *IEEE Transactions on Power Systems*, vol. 33, no. 6, pp. 6667–6674, November 2018.

[4] M. A. Baferani, M. R. Chalaki, N. Fahimi, A. A. Shayegani and K. Niayesh, "A novel arrangement for improving three phase saturated-core fault current limiter (SCFCL)," *IEEE Texas Power and Energy Conference (TPEC)*, pp. 1-6, 2018.

[5] J. Verboomen, D. V. Hertem, P. H. Schavemaker, W. L. Kling, and R. Belmans, "Border-flow control by means of phase shifting trans-formers," *IEEE Lausanne Power Tech*, pp. 1338–1343, 2007.

[6] P. Kundur, "Power system stability and control," McGraw-Hill Professional, 1994.

[7] M. H. R. Koochi, S. Esmaeili, and P. Dehghanian, "Coherency detection and network partitioning supported by wide area measurement system," *IEEE Texas Power and Energy Conference (TPEC)*, pp. 1–6, 2018.

[8] J. Verboomen, D. V. Hertem, P. H. Schavemaker, W. L. Kling, and R. Belmans, "Phase shifting transformers: principles and applications," *International Conference on Future Power Systems*, 2005.

[9] J. M. Cano, M. R. R. Mojumdar, J. G. Nomiella, and G. A. Orcajo, "Phase shifting transformer model for direct approach power flow studies," *International Journal of Electrical Power and Energy Systems*, vol. 91, pp. 71–79, 2017.

[10] U. Khan, "Modeling and protection of phase shifting transformers," Ph. D. dissertation, University of Western Ontario, 2013. [Online]. Available: https://ir.lib.uwo.ca/etd/1701/

[11] A. S. Siddiqui, S. Khan, S. Ahsan, M. I. Khan and Annamalai, "Application of phase shifting transformer in Indian Network," *International Conference on Green Technologies (ICGT)*, Trivandrum, 2012, pp. 186-191,

[12] T. J. Morrell and J. G. Eggebraaten, "Applications for phase-shifting transformers in rural power systems," *IEEE Rural Electric Power Conference (REPC)*, 2019.

[13] A. Baczynska and W. Niewiadomski, "The impact of the phase-shifting transformers on the power flow in the transmission grid," pp. 1–5, June2018.

[14] R. D. Youssef, "Phase-shifting transformers in load flow and short-circuit analysis: modelling and control," *IEEE Proceedings C - Generation, Transmission and Distribution*, vol. 190, no. 4, pp. 331–336, July 1993.

[15] A. Baczyńska and W. Niewiadomski, "The Impact of the Phase-Shifting Transformers on the Power Flow in the Transmission Grid," *15th International Conference on the European Energy Market (EEM)*, Lodz, 2018, pp. 1-5,

[16] Y. Kawaura, S. Yamanouchi, M. Ichihara, S. Iwamoto, Y. Suetsugu, and T. Higashitani, "Phase-shifting transformer application to power-flow adjustment for large-scale pv penetration," *IEEE Region 10 Conference (TENCON)*, pp. 3328–3331, 2016.

[17] J. Cano, R. Mojumdar, J. Norniella, and G. Orcajo, "Phase shifting transformer model for direct approach power flow studies," *International Journal of Electrical Power and Energy Systems*, vol. 91, pp. 71-79, 2017.

[18] A. El Hraïech, K. Ben-Kilani and M. Elleuch, "Control of parallel EHV interconnection lines using Phase Shifting Transformers," *IEEE 11th International Multi-Conference on Systems, Signals & Devices (SSD14)*, Barcelona, pp. 1-7, 2014.

[19] J. Yeo, P. Dehghanian, and T. J. Overbye, "Power flow consideration of impedance correction for phase shifting transformers," *IEEE Texas Power and Energy Conference (TPEC)*, pp. 1–6, February 2019.

[20] I. Kasikci, "Short circuits in power systems: A practical guide to IEC 60909-0," Wiley-VCH, second edition, 2018.

[21] J. Glover, T. Overbye, and M. Sarma, Power System Analysis and Design. Cengage Learning, 2016.

[22] A. B. Birchfield, T. Xu, K. M. Gegner, K. S. Shetye and T. J. Overbye, "Grid Structural Characteristics as Validation Criteria for Synthetic Networks," *IEEE Transactions on Power Systems*, vol. 32, no. 4, pp. 3258-3265, July 2017.

[23] Electric Grid Test Case Repository, [Online]. Available: https://electricgrids.engr.tamu.edu/electric-grid-test-cases/

[24] Eastern Interconnection Reliability Assessment Group, "Multiregional modeling working group," Procedural manual, version 25, March 2020.

[25] J. D. Weber and T. J. Overbye, "Voltage contours for power system visualization," *IEEE Trans. on Power Systems*, pp. 404-409, 2000.

[26] T. J. Overbye, J. D. Weber and K. J. Patten, "analysis and visualization of market power in electric power systems," *Proc. 32nd Hawaii International Conference on system Sciences*, Maui, HI, January 1999.

[27] T. Satoh, H. Tanaka and S. Iwamoto, "ATC Improvement by Phase Shifter Application considering Dynamic Rating," *39th North American Power Symposium*, Las Cruces, NM, 2007, pp. 528-533,